\documentclass[aps,prl,twocolumn,superscriptaddress,longbibliography]{revtex4-2}
\usepackage{amsmath, physics,amssymb}
\usepackage[T1]{fontenc}
\usepackage{amsmath,amssymb,mathtools}
\usepackage{amsthm}
\usepackage{graphicx}
\usepackage{booktabs}
\usepackage{xcolor}
\usepackage{microtype}

\usepackage[colorlinks=true,linkcolor=blue,citecolor=blue,urlcolor=blue]{hyperref}
\usepackage[capitalise,nameinlink,noabbrev]{cleveref}

\graphicspath{{figures/}}
\setcitestyle{sort&compress,numbers}

\theoremstyle{definition}

\theoremstyle{remark}

\begin{document}

\title{Nonlocality relates cuprate photoemission $\omega/T$ scaling to a broad dielectric loss peak}

\author{Jian Xian Sim}
\email{simjianxian@u.nus.edu}
\affiliation{Centre for Quantum Technologies, National University of Singapore, 3 Science Drive 2, Singapore 117543}
\affiliation{Summer Visitor: Yukawa Institute for Theoretical Physics, Kyoto University, Kyoto 606-8502, Japan}

\date{\today}

\begin{abstract}
 Photoemission spectroscopy on high $T_c$ cuprates reveals a puzzling nodal self-energy with $\omega/T$ scaling and an exponent varying continuously with doping. Meanwhile, electron energy loss spectroscopy finds a broad $\sim 1$eV peak whose physical origin remains debated. We propose they share a common origin by introducing a phenomenological model of nonlocal screening in a many-body charged fluid. The model's key feature is spatially nonlocal effective repulsions between carriers $V_{\alpha}(r) \sim 1/r^\alpha$, with doping-dependent $\alpha$. The model predicts a falsifiable relationship between independently measured self-energy exponent $\gamma$ and loss function peak dispersion exponent $\alpha$, as $\gamma = 2-\frac{1}{\alpha}$. This can be tested using modern ARPES and EELS. Nonlocality provides an alternative route to $\omega/T$ scaling in observables, although it is conventionally regarded as a signature of quantum criticality.
\end{abstract}
\maketitle

\paragraph*{Introduction.} A clear grasp of high temperature superconducting materials such as cuprates remain elusive despite robust empirical scaling laws~\cite{Hartnoll_2022, Keimer2015}. Of great interest at temperatures above the superconducting condensation, is the anomalous normal state often referred to as a strange metal. 

Existing studies of cuprates place significant focus on the strong local correlations, captured by paradigmatic models such as the two-dimensional Hubbard or Emery model with strong on-site repulsion between electrons~\cite{Emery_1987}. Due to immense difficulties in performing analytics or classical simulations for such models, quantum simulations of local models have been pursued~\cite{lange2026realizingemerymodeloptical}. In similar fashion, analyses of strange-metal dynamics in large part assume effectively
local interactions and diffusive hydrodynamics~\cite{Hartnoll_2022, Gu_2017, Zaanen_2024, Patel_2018}. 

However, doping a Mott or charge transfer insulator~\cite{Zaanen1985} might lead to incomplete screening, unable to restore effectively local interactions. Furthermore,  for weak interlayer tunnelling, each copper oxide conducting plane is effectively two-dimensional. This causes poorer screening than in three dimensions, even in an ideal Fermi liquid. 

Long-ranged poorly screened effective repulsions between carriers
$V_{\alpha}(r)\sim 1/r^\alpha$, where $\alpha<2$ for a two-dimensional plane, can qualitatively alter many-body dynamics. In particular, sufficiently long-range interactions invalidate conventional Lieb–Robinson bounds on information propagation and can undermine ordinary diffusive hydrodynamics~\cite{LiebRobinson1972,Kuwahara_2020,Kuwahara_2021, Nishikawa_2025,Nishikawa2026Microscopic}~\footnote{To intuitively see why the system loses its intrinsic energy scale, one can consider integrating the potential $\int d^Dr \ \frac{1}{r^\alpha}$ and observe if $\alpha < D$, the integral diverges due to large $r$. This is the main reason why rigorous bounds on both information propagation and diffusive hydrodynamics fail.}. This suggests a route to anomalous dynamical scaling rooted in spatial nonlocality, rather than quantum criticality. 

Motivated by this observation, we investigate whether nonlocal screening can account for the anomalous photoemission and dielectric loss spectra of cuprates. We develop a falsifiable phenomenology using angle-resolved photoemission spectroscopy (ARPES) and momentum-resolved electron energy loss spectroscopy (EELS), through a case study of Bi2201 and Bi2212 cuprates.

Our approach follows the phenomenological philosophy pioneered by Landau's theory of Fermi liquids~\cite{Landau1957, Landau1959}: rather than committing to a microscopic Hamiltonian, we introduce a small number of physical assumptions and derive quantitative relations between experimentally accessible observables. The predictive value of such a theory lies in constraining independent measurements through common parameters.

In this work's model of cuprates, we first motivate the assumption that dynamical screening is formulated through nonlocal hydrodynamics in Eqn~\ref{eqn:dynamicalscreening}. This formulation enables calculation of the self-energy and loss function within this effective theory. Key analytical results are (I) $\omega/T$ scaling of the imaginary self-energy with exponent $\gamma = 2-\frac{1}{\alpha}$, valid for $1 < \alpha < 2$ in Eqn~\ref{eqn:scalecovariant}. (II) A broad dielectric loss peak with a fractional dispersion exponent $\alpha$, in Eqn~\ref{eqn:lossfn}. 

At this point we note that $\omega/T$ scaling of observables is widely viewed as diagnostic of quantum criticality and its associated scale-invariance~\cite{Hartnoll_2022, Zaanen_2024, SachdevKeimer2011, Helton_2010}. However, our result (I) arises from nonlocality.

\begin{figure}
    \centering
    \includegraphics[width=1\linewidth]{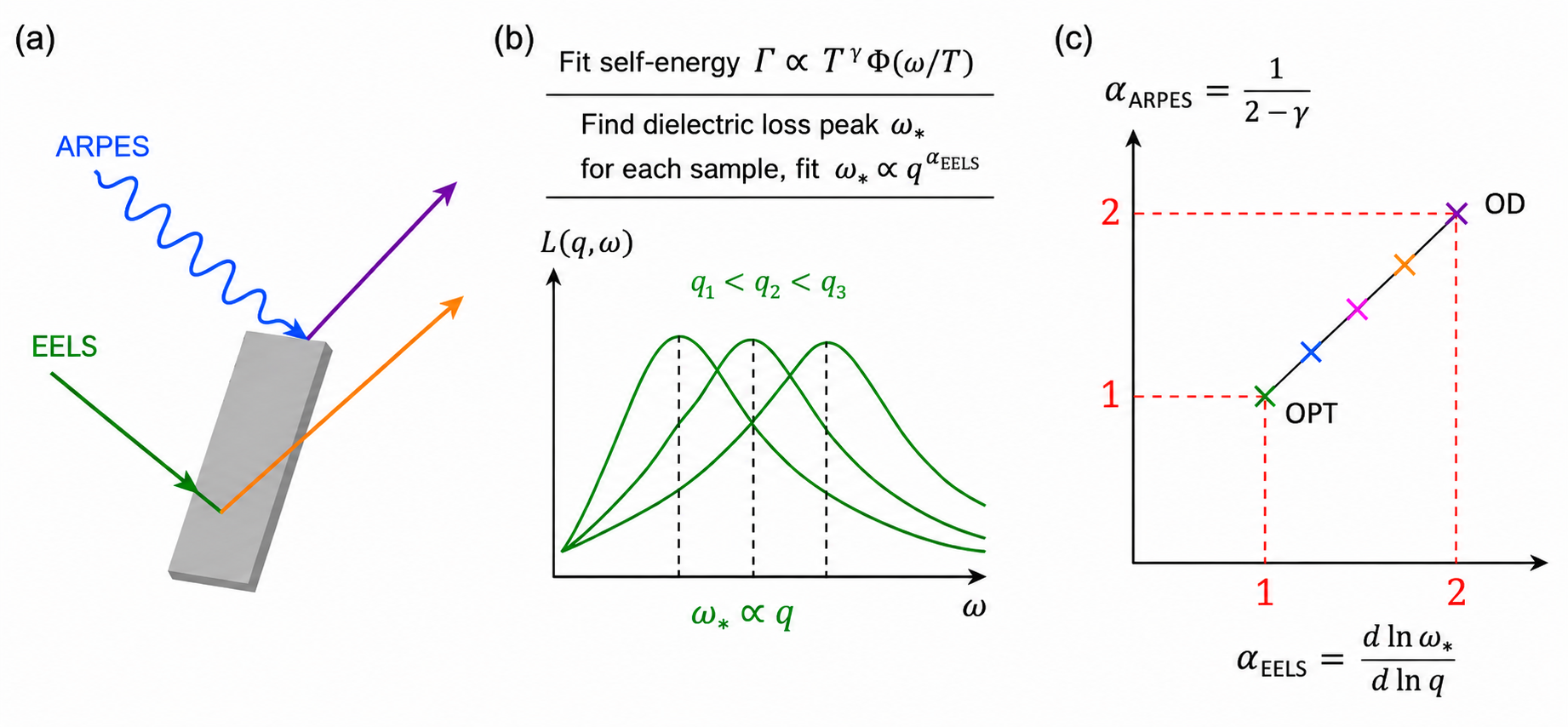}
    \caption{Experimental falsification of nonlocality: (a) For each cuprate sample with doping varying from optimal to overdoped, conduct ARPES (photon in, electron out) and EELS (electron in, electron out) measurements. (b) Fit ARPES self energy $\Gamma(\omega,T)$ data to find $\alpha_{ARPES}$. Fit EELS dielectric loss $L(q,\omega)$ drifting peak data at small $q$ in fine steps, to find $\alpha_{EELS}$. E.G. in the representative plot at optimal doping, $\omega_* \propto q$ with $\alpha = 1$. (c) Our model predicts Eqn~\ref{eqn:falsify}, cross-spectroscopy consistency $\alpha_{ARPES} = \alpha_{EELS}$ for each sample (schematic data points). `OPT' in green means optimal doping, `OD' in purple means large overdoping.}
    \label{fig:falsifyexpt}
\end{figure}

From (I) and (II) a cross-spectroscopy relation Eqn~\ref{eqn:falsify} arises between ARPES and EELS. This provides a direct route to experimental falsification (see Fig~\ref{fig:falsifyexpt}), as we now outline.

At an intuitive level, more doping should enhance screening between carriers, thus shortening the effective interaction range and increasing $\alpha$. This is mostly consistent with existing photoemission experiments in the optimal to overdoped regime~\cite{Smit_2024,reber2015powerlawliquid}. We also discuss relevant limits, such as near optimal doping where our theory's self-energy approaches linear scaling laws in energy $\omega$ and temperature $T$, characteristic of a Marginal Fermi liquid~\cite{Varma_1989}.  On the other hand, momentum-resolved EELS peak dispersion has not been experimentally explored at sufficiently small momentum. Until the decisive cross-spectroscopy test is performed, the relationship between the anomalous self-energy and broad dielectric loss peak may remain unclear.

We have a targeted reason for choosing ARPES and EELS for falsification, which is that they retain momentum and frequency information intrinsic to the many-body electron-electron dynamics. Such intrinsic many-body information can help formulate smoking gun falsification experiments. 

Historically, transport scaling laws with no momentum or frequency resolution, such as robust $T$-linear resistivity at optimal doping, were more practical to measure. Given the striking deviation from $T^2$ Fermi liquid scaling, resistivity and other transport anomalies have garnered significant attention~\cite{Varma_1989, Hartnoll_2015, Hartnoll_2022, Phillips_2022}. However, to calculate transport quantities can involve many intermediate steps and approximations, and may even require introducing objects beyond electron-electron interactions. For example, electrical current cannot be relaxed unless momentum conservation is broken, one avenue being Umklapp scattering of carriers with the ionic lattice~\cite{Hartnoll_2022, Zaanen_2024}. 

As a result, competing theoretical proposals can produce similar observable scalings, difficult to distinguish experimentally. Furthermore, from a conceptual viewpoint, it becomes harder to identify amongst various approximations what the physical core of a given theoretical framework is. 

Fortunately, modern advances in ARPES and EELS~\cite{Liu2008LaserARPES, Zhu2015HREELS} make possible rigorous experimental tests of theoretical proposals demanding high momentum and frequency-resolution, such as the model developed in this work.

\paragraph*{Brief recap of ARPES on Bi2212 and Bi2201} Nodal ARPES on Bi2212 were fitted to a `power-law liquid (PLL)' form by Reber et al.~\cite{reber2015powerlawliquid}, continuously interpolating between underdoped and overdoped samples $p \sim 0.09 -0.21$ through a single exponent $\beta(p) \sim 0.34 - 0.62$, as
\begin{equation}\label{eqn:nodalselfenergy}
    \Gamma \equiv -Im[\Sigma(\omega, T)] = \Gamma_0(\beta) + \lambda T^{2\beta} \frac{[(\omega/T)^2 + (Ck_B)^2]^\beta}{(\omega_N)^{2\beta-1}},
\end{equation}
inelastic scattering $\Gamma_{inel}$ for the second term, note the function displays $\omega/T$ scaling with dimensionless function $\Phi_{PLL} = \frac{[(\omega/T)^2 + (Ck_B)^2]^\beta}{(\omega_N)^{2\beta-1}}$. This is similar to our result setting $\gamma = 2\beta$ (see Eqn~\ref{eqn:scalecovariant}), the important shared property is that both can be written simply as $\propto T^\gamma \Phi(\omega/T)$. Around optimal doping, fixing $\beta = 0.5$ one has a Marginal Fermi liquid~\cite{Varma_1989}. Newer experiments on Bi2201~\cite{Smit_2024} fit the slightly underdoped $p \sim 0.14$ to very overdoped regime $p \sim 0.29$ as $\beta \sim 0.52 - 0.84$.

However, there are two more parameters in Reber et al.'s Eqn~\ref{eqn:nodalselfenergy} than our theory's Eqn~\ref{eqn:scalecovariant}. Our dimensionless $\omega/T$ scaling function $\Phi_\gamma(\omega/T)$ arises from counting one-loop hydrodynamic scattering processes, and contains no extra fitting parameters. Thus, our theory is more phenomenologically constrained when fitted to ARPES data. 

\paragraph*{Model Setup: Screening, nonlocality, hydrodynamics} We do not fix an explicit microscopic Hamiltonian. We first consider screening of Coulomb repulsions between charges in a $2D$ plane. Observe that an undoped Mott insulator can only screen by local electronic orbital distortions changing its dielectric constant $\epsilon_0 \rightarrow \epsilon$, but total absence of mobile charges keeps the global decay law at $1/r$. On the other hand, the idealized overdoped Fermi liquid in $2D$ with many mobile charges, has a modified global decay law $1/r^3$ by Thomas-Fermi theory. For doping levels between these limits, we find it reasonable to postulate between charges on sites $i,j$, a screened effective repulsion $V_{\alpha}(\mathbf{r_i}-\mathbf{r_j}) = g_{\alpha}/|\mathbf{r_i}-\mathbf{r_j}|^\alpha$, for $1 \le \alpha \le 3$, $g_\alpha$ a coupling constant. 

We expect that $\alpha$ generally increases with doping, as mobile charge carriers restore good screening. 
We will not deduce from microscopics how $\alpha$ varies with doping $p$. Rather, we make several microscopically-agnostic calculations and constrain the evolution of $\alpha(p)$ from ARPES data, then use $\alpha$ to predict EELS results not yet measured.

From the above postulate, now coarse-grain the lattice picture into a continuum, consider a charged fluid model where $n(\mathbf{r},t)$ is the charge density at each point in space and time. Then in momentum variable $\mathbf{q}$ (Fourier transform of $\mathbf{r}$), the many body system has free energy functional $F[n] \sim \int_{\mathbf{q}} \ n(-\mathbf{q}) V(\mathbf{q})n(\mathbf{q})$, where $\int_{\mathbf{q}} \equiv [1/(2\pi)^2] \int d^2q$. At leading order near $\mathbf{q} \rightarrow 0$, $V(\mathbf{q}) \sim \frac{1}{|\mathbf{q}|^{2-\alpha}}$ when $\alpha < 2$, but for $\alpha > 2$ $V(\mathbf{q}) \sim V(0) \sim O(1)$ since the Fourier Transform converges. Crucially, $V(\mathbf{q})$ cannot be written down as a power series in $|\mathbf{q}|^2$ when $\alpha<2$. The basic assumption underlying the conventional free energy local gradient expansion fails here at leading order~\footnote{This is a drastic breakdown of the gradient expansion to be distinguished from Pippard's example~\cite{Pippard1953, Phillips_2022} which is nonlocality at subleading order. Also, note that $2<\alpha<4$ has a nonanalytic subleading term $\sim |\mathbf{q}|^{\alpha-2} \gg |\mathbf{q}|^{2}$}. This nonlocal free energy will influence charge relaxation as follows. 

Consider a non-rigorous approach for a current density $\mathbf{j}(\mathbf{r},t)$, by assuming the (i) continuity law $\nabla \cdot \mathbf{j} + \partial_t n = 0$, (ii) chemical potential gradient law $\mathbf{j} = -\sigma\nabla\mu $ where $\mu(\mathbf{q}) \equiv \frac{\delta F}{\delta n(-\mathbf{q})}$ is a functional derivative. Applying a Fourier Transform in frequency $\omega$, we obtain hydrodynamic poles as (A) $\alpha > 2$ diffusive, $\omega = -iD_\alpha|\mathbf{q}|^2$ and (B) $\alpha < 2$ superdiffusive, $\omega = -iD_\alpha|\mathbf{q}|^\alpha$.~\footnote{Our heuristic argument can fail in one dimension as shown by Nishikawa and Saito~\cite{Nishikawa_2025, Nishikawa2026Microscopic}. But their results morally support our case, because in $D=1$, superdiffusion occurs for $\alpha < 3/2$ showing the ubiquity of superdiffusion. Furthermore, their diffusion threshold in $D=2$ is precisely $\alpha = 2$, along with more direct analysis by Kiselev for charged fluids in two dimensions corroborating our heuristic argument here~\cite{Kiselev_2021}.}

% From the above observations, we postulate the proper polarizability of the high $T_c$ strange metal medium as 
% \begin{equation}
%     \Pi^R_{eff}(q,\omega) =  \Pi^R_{prop}(q,\omega) + [V_{\alpha}^{-1}(q)-V^{-1}_{bare}(q)]h^{R}(\frac{\omega}{\Omega_{UV}})
% \end{equation}
% where $V_{bare}(q)$ is the bare Coulomb potential, $f^R (\omega/\Omega_{UV})$ is a phenomenological function such that $(\omega/\Omega_{UV}) \ll 1$ gives $h^R \rightarrow 1$ but $(\omega/\Omega_{UV}) \gg 1$ gives $h^R \rightarrow 0$. The second term is then thought of as an anomalous spatially nonlocal proper polarization over an intermediate energy and momentum window, determined by hydrodynamics. 

% We can interpret our hypothesis on proper polarizability as assertion of the central role played by mesoscopic energy scales~\cite{laughlin2000middle}, for capturing the physical essence of the many-body electron medium in high $T_c$ materials. This is analogous to Pines and Bohm asserting the random phase approximation in their analysis of conventional metals~\cite{PhysRev.85.338}~\footnote{RPA assumes that the proper polarizability of the interacting system can be approximated with that of the non-interacting system.}.

In light of the above, before the nonlocal interaction is included, we have a diffusive irreducible polarization $\Pi^R_{irr}(q, \omega) = \chi\frac{Dq^2}{Dq^2-i \omega}$, $\chi \equiv (\frac{\partial n}{\partial \mu})_T$ the compressibility~\footnote{This form follows directly from charge conservation, hydrodynamics, and linear response, for example see~\cite{RevModPhys.57.287}. It is equivalent to say that the measurable polarizability $\chi^{R}_{nn}(q,\omega) = (\sigma q^2)/(D_\alpha q^\alpha + D q^2 - i\omega)$}. Apply $(W^{R})^{-1} = V_{\alpha}^{-1} + \Pi^R_{irr}$ to get 
\begin{equation}\label{eqn:dynamicalscreening}
    W^{R}(\mathbf{q},\omega) = V_\alpha(\mathbf{q})
\frac{D|\mathbf{q}|^2-i\omega}
{D|\mathbf{q}|^2[1+\chi V_\alpha(\mathbf{q})]-i\omega},
\
D \equiv \frac{\sigma}{\chi},
\end{equation} Strictly speaking, in the limit of $\omega \rightarrow \infty$ at fixed $|\mathbf{q}|$, the bare Coulomb law $W(q,\omega) \sim \frac{1}{|\mathbf{q}|}$ should be restored. Thus, our formula for $W$ can only hold below $ \omega \ll \Omega_{UV}$, a cutoff scale~\footnote{Furthermore, $\lim_{|\mathbf{q}|\rightarrow 0} \lim_{\omega\rightarrow 0}W(\mathbf{q},\omega) = 1/\chi$, conventional static screening is restored.}. 

For a clearer picture, consider the material's charge relaxation in real space. It seems unlikely superdiffusion occurs at macroscopic scales across the entire sample. Above some material-dependent length scale, conventional diffusion may be restored. On the other hand, hydrodynamics is also unlikely to arise at small distances on the order of lattice constants. Therefore, the superdiffusion occurring in this theory occupies a rather unsettling mesoscopic regime~\cite{laughlin2000middle}. The only way to determine whether it makes sense as an effective theory of a given material, is to experimentally test the observable consequences we now develop.

% We define $D_\alpha \equiv D\chi g_{\alpha}, \ D_2 \equiv D(1+\chi V_0)$ as generalized diffusion coefficients and $V_0 \equiv\lim_{q\rightarrow 0}V_\alpha(q)$.
% \begin{align}
% W^R(\mathbf q,\omega)
% &=
% W_{\mathrm{stat}}(\mathbf q)
% +
% \left[
% V_\alpha(\mathbf q)-W_{\mathrm{stat}}(\mathbf q)
% \right]
% \frac{-i\omega}{\omega_{\mathbf q}-i\omega},
% \label{eq:screening_interpolation}
% \\
% W_{\mathrm{stat}}(\mathbf q)
% &\equiv
% \frac{V_\alpha(\mathbf q)}
% {1+\chi V_\alpha(\mathbf q)},
% \\
% \omega_{\mathbf q}
% &\equiv
% D|\mathbf q|^2
% \left[
% 1+\chi V_\alpha(\mathbf q)
% \right].
% \end{align}
% Thus, \(W^R\) interpolates from the statically screened interaction
% \(W_{\mathrm{stat}}\) at low frequency to the unscreened effective
% interaction \(V_\alpha\) at high frequency, with relaxation rate
% \(\Gamma_{\mathbf q}\). 
Define $D_\alpha \equiv D\chi g_{\alpha} = \sigma g_\alpha$. Note $\chi V_\alpha(\mathbf{q}) \gg 1$ for small $|\mathbf{q}|$, not literally taking $|\mathbf{q}| \rightarrow 0$. This assumption can be violated at very low doping near the Mott insulator due to low $\chi$. We get the dissipative part with a simple decomposition
\begin{align}
    -\operatorname{Im} W^{R}(\mathbf{q},\omega) 
    &\approx g_\alpha D_\alpha |\mathbf{q}|^{2\alpha -2 } \frac{\omega}{\omega^2 + (D_\alpha |\mathbf{q}|^\alpha)^2} \nonumber \\
    &\equiv \omega^{1-\frac{2}{\alpha}} h\!\left(\frac{\omega}{|\mathbf{q}|^\alpha}\right), \label{eqn:imaginaryscreening}
\end{align}
where $h(u) = g_\alpha D_\alpha \frac{u^{2/\alpha}}{u^2 + D_\alpha^2}$. If $\alpha > 1$, the \textit{characteristic momentum} maximising $-ImW^R$  is $q_{\omega} \sim (\omega/D_\alpha)^{1/\alpha}$.

\paragraph*{Quasiparticle decay with $\omega/T$ scaling}
% \JX{The retarded fermionic Green's function characterizes the extent to which a fermionic charge carrier's `particle-ness' is preserved under many-body dynamics, defined as
% \begin{equation}
%     G^R(r,t) = -i\langle \{ c(r,t), c^\dagger(0,0) \}\rangle
% \end{equation}
% where the average is taken over the relevant density matrix. }

 The model's physical picture for the single-fermion self-energy is a charge carrier propagating in a hydrodynamic medium in thermal equilibrium at temperature $T$. The movement of a given carrier causes density fluctuations of the surrounding charged fluid. By scattering off such density relaxation modes, the carrier's information disperses and it `loses memory of itself'. 
 
 This process is formally described by the retarded single fermion Green's function $G^R(\mathbf{r},t;T) \equiv -i \Theta(t) \langle \{ c(\mathbf{r},t), c^\dagger(0,0)\} \rangle_{T}$ with expectation value evaluated at the Gibbs state, $\Theta(t)$ the step function. When the system has a quasiparticle description we write 
 \begin{equation}
     G^R(\mathbf{k},\omega;T) = \frac{1}{\omega - \xi_{\mathbf{k}}  - \Sigma(\mathbf{k}, \omega)},
 \end{equation}
where $\xi_{\mathbf{k}}$ is energy difference from Fermi surface. In our setup, the free fermion Green's function $G_0(\mathbf{k},\omega) \equiv 1/(\omega - \xi_\mathbf{k} + i0^+ )$ is dressed by the interaction propagator $W$, whose pole corresponds to hydrodynamic modes $\sim 1/(-i\omega +D_\alpha |\mathbf{q}|^\alpha)$~\footnote{It must be emphasized that such hydrodynamic modes do not constitute particle-like excitations. For contrast, a phonon mode has propagator $\sim \frac{1}{\omega^2 - c^2q^2}$ with a real pole. On the other hand, our decaying hydrodynamic mode has propagator $\sim \frac{1}{-i\omega + D_\alpha q^\alpha}$, an imaginary pole.}, giving a self-energy $\Sigma({\mathbf{k}},\omega ; T) = G_0^{-1}({\mathbf{k}},\omega) - G^{-1}({\mathbf{k}}, \omega ; T)$. 

We choose a single smooth Fermi surface patch for the scattering near one selected momentum $\mathbf{k}_F$, define the Fermi velocity $v_F \equiv |\mathbf{v}_F| \equiv |\nabla_{\mathbf{k}} \xi_{\mathbf{k}}|_{\mathbf{k}_F}$ perpendicular to the Fermi surface. Let the quasiparticle decay rate from inelastic scattering be $\Gamma_{inel}(\mathbf{k}_F,\omega, T) \equiv -\operatorname{Im}\Sigma^{R}(\mathbf{k}_{F},\omega;T)$. We calculate the one-loop fermionic self-energy using the $G_0 W$ approximation, summing four elementary electron/hole emission and absorption processes involving a single hydrodynamic density relaxation mode of momenta $\mathbf{q}$, energy $\Omega$ (see Supplemental Material). With standard approximations (A) \textit{Small-Angle Scattering $q \ll k_F$} and (B) \textit{Tangential Momentum Dominance $|\mathbf{q}| \approx |q_{\parallel}|$} where scattering is dominantly parallel to Fermi surface, elaborated in the Supplemental Material, for $\gamma , \alpha > 1$ we get
\begin{align}\label{eqn:inelscattering}
\Gamma_{inel}(\mathbf{k}_F,\omega, T) 
&=
\int_{\mathbf{q}}
\int_{-\infty}^{\infty} d\Omega\,
\bigl[-\operatorname{Im}W^{R}(\mathbf{q},\Omega)\bigr]
\nonumber\\
&\hspace{-3.5em}\times
\left[
n_{B}(\Omega)
+
n_{F}\!\left(-\xi_{\mathbf{k}_{F}-\mathbf{q}}\right)
\right]
\delta\!\left(
\omega-\Omega-\xi_{\mathbf{k}_{F}-\mathbf{q}}
\right)
\\
&\approx
A_\gamma T^\gamma
\Phi_\gamma\!\left(\frac{\omega}{T}\right).
\label{eqn:scalecovariant}
\end{align}
where $\gamma = {2-\frac{1}{\alpha}}$, $\Phi_{\gamma}(x) \equiv \int^{\infty}_0 dy \ y^{\gamma-1}[\frac{2}{e^{y}-1} + \frac{1}{e^{y-x}+1} + \frac{1}{e^{y+x}+1}]$ is a dimensionless function, $A_\gamma = (g_\alpha D_\alpha^{1-\gamma})/[4\pi \alpha v_F \sin(\pi/2\alpha)]$ a constant with $[A] = E^{1-\gamma}$~\footnote{So that units are right, $[\Gamma] = [\omega] = [T] = E$.}. $n_B, n_F$ are Bose and Fermi-Dirac distributions. 

Note $\gamma$ decreases together with $\alpha$, physically it means longer-ranged effective interactions enhance scattering rates and shorten quasiparticle lifetimes. 
\begin{figure*}[t]
    \centering
    \includegraphics[width=1\linewidth]{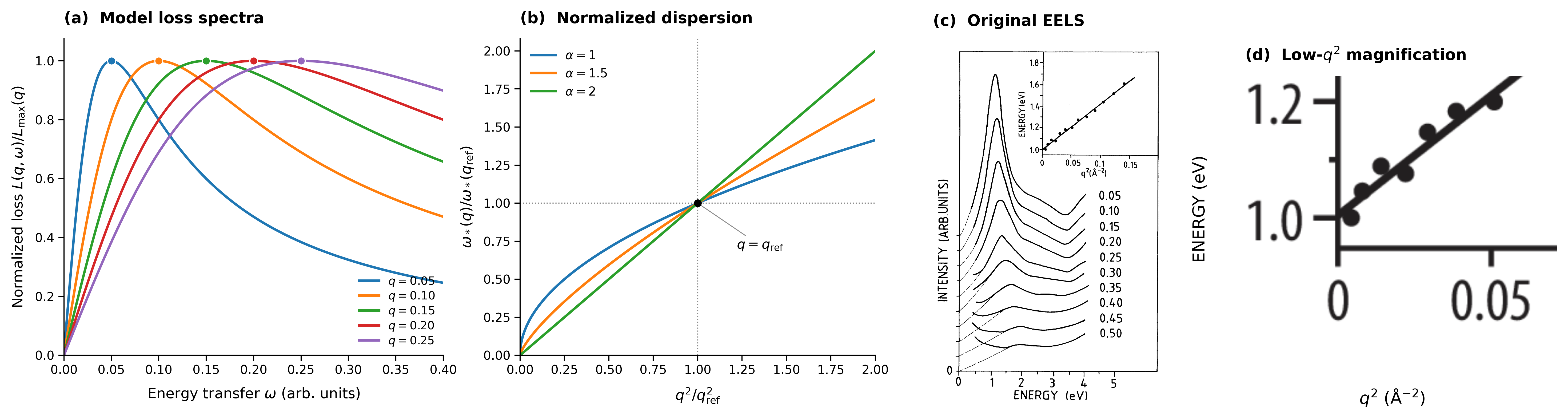}
    \caption{General guide: Compare Panel (a) to (c) and (b) to (d). Panel (a) is our two-dimensional model's dielectric loss function Eqn~\ref{eqn:lossfn}, the peak labelled with a dot for $\alpha = 1$, plotted for momentum values $q=0.05 -0.25$ in arbitrary units. The layered model's loss function curve has no qualitative difference in shape, only a shift in the $x$-axis intercept. Panel (b) plots the peak dispersion relations from our model with arbitrary reference scale $q_{ref}$ for $\alpha = 1,1.5,2$, with $q^2$ on the horizontal axis demonstrating the concave-down behaviour. Panel (c) and (d) are transmission-EELS data on Bi2212 from Nucker et al.~\cite{Nucker1989} focusing on a broad $1$eV peak's dispersion, with permission from Jorg Fink, Copyright license by the American Physical Society. Panel (c) shows over large momentum regimes a $q^2$-like dispersion. In Panel (d), we magnify the inset of Panel (c) to retain only values $q \lesssim 0.22 \text{\AA}^{-1}$. In the regime, when a $q \ll d^{-1} \sim 0.3 \text{\AA}^{-1}$ expansion starts to be valid, experimentally there remains a possibility that the dispersion exponent is not $2$ and may be concave down, c.f. Panel (b). Experimental resolution forty years ago~\cite{Nucker1989} was too low to make any conclusion, motivating us to revisit this question today.}
    \label{fig:EELS}
\end{figure*}
The model does not determine whether the prefactor $A_\gamma$ is temperature-dependent, this is decided by experimental fit. If $A_\gamma$ is independent of temperature, then $\Gamma_{inel}$ has strict scale-covariance with scaling dimension $\gamma$, defined as $\Gamma_{inel}(b \omega, b T) = b^\gamma \Gamma_{inel}(\omega,T)$.

The procedure to calculate Eqn~\ref{eqn:scalecovariant} is standard and similar to that of electron self-energy from electron-phonon coupling~\cite{zhou2006angleresolvedphotoemissionspectroscopyelectronic}, only differing in the `scattering spectrum'. In electron-phonon scattering, it is the Eliashberg spectrum $\alpha^2_{k}F(\Omega)$. The main novelty of our electron-hydrodynamic scattering process, is a superdiffusive charge density relaxation spectrum $\propto\Omega^{\gamma-1}$ (compare Eqn (4.7) of Zhou et al.~\cite{zhou2006angleresolvedphotoemissionspectroscopyelectronic} to Eqn (20) of our Supplemental Material). 

Some comments on our self-energy result:

(i) To see why the self-energy exponent is $2-\frac{1}{\alpha}$, consider the following dimensional argument. The small-angle scattering condition along with $|\mathbf{q}| \approx q_{\parallel}$, roughly speaking, restricts the scattering to a thin window around the Fermi surface, reducing the integral to $\int d\Omega \int dq_{\parallel} \bigl[-\operatorname{Im}W^{R}(q_{\parallel},\Omega)\bigr] \sim \int d\Omega [\int dq_{\parallel} \ \Omega^{1-\frac{2}{\alpha}} h(\frac{\Omega}{q_{\parallel}^\alpha})]$. The \textit{characteristic momentum} $ q_{\parallel} \sim \Omega^{1/\alpha}$ sets the scale of dominant momentum contributions, $[\int dq_{\parallel} \ \Omega^{1-\frac{2}{\alpha}} h(\frac{\Omega}{q_{\parallel}^\alpha})] \sim \Omega^{1-\frac{1}{\alpha}}$.
Integrating energy transfer $\int d\Omega\,\Omega^{1-\frac{1}{\alpha}}\sim\Omega^{2-\frac{1}{\alpha}}$ gives energy scaling dimension $\gamma = 2-\frac{1}{\alpha}$.

(ii) Letting $\alpha \rightarrow 1^+$ in Eqn~\ref{eqn:scalecovariant}, $\gamma \rightarrow 1^+$ with shortening quasiparticle lifetime, approaching the Marginal Fermi liquid self-energy. Thus, by comparing with experiment~\cite{reber2015powerlawliquid, Smit_2024}, our result indicates that the effective repulsion has $\sim 1/r$ decay law around optimal doping, specifically $p \sim 0.16-0.17$ for Reber et al~\cite{reber2015powerlawliquid}. Omitting $T$-dependence for analytical simplicity, at $\gamma = 1$ we obtain exactly the Marginal Fermi liquid self-energy (see Supplemental Material). 

(iii) Experimentally, increasing doping raises the fitted $\gamma$~\cite{reber2015powerlawliquid, Smit_2024}, corresponding in our phenomenology to increasing $\alpha$, consistent with progressively improved screening.

(iv) When $\alpha > 2$, diffusive hydrodynamics is restored, now $\gamma = 1.5$ is fixed. The largest experimentally fitted exponent $\gamma_{expt} \sim 1.68$ in~\cite{Smit_2024}, at $p \sim 0.29$ outside the superconducting dome, exceeding the diffusive limit of our result. This discrepancy deserves closer experimental scrutiny and may indicate limitations of our theory at large overdoping~\footnote{However, their measured growth of $\gamma$ with doping tapers off in the overdoped regime.}.

In summary, the ARPES results in optimal to overdoped regimes within the superconducting dome closely relate to $\gamma > 1$ of Eqn~\ref{eqn:inelscattering}.

\paragraph*{Underdoped regime.}
In the underdoped regime, Reber et al. reported nodal ARPES exponents as small as $\gamma \sim 0.68$ for Bi2212, beyond the validity of our present approximation. In contrast, Smit et al. reported $\gamma \gtrsim 1.04$ even for mildly underdoped Bi2201. Resolving this experimental difference would be valuable. A quantitative description of more underdoped cuprates likely lies outside the model's present scope.

\paragraph*{Falsification using loss spectroscopy} 
The prediction made here of dielectric loss function properties and its relation to the self-energy, is to my knowledge, unique to the hydrodynamic screening Eqn~\ref{eqn:dynamicalscreening}. First, define the dielectric function $\epsilon(\mathbf{q},\omega)$, from $W^R(\mathbf{q},\omega) = \frac{V_C(\mathbf{q})}{\epsilon(\mathbf{q},\omega)}$ with $V_C = (2\pi e^2/q)$ now the bare Coulomb potential in a $2D$ plane. Similar to Eqn~\ref{eqn:imaginaryscreening}, a direct calculation gives the loss function
\begin{equation}\label{eqn:lossfn}
\begin{aligned}
L(\mathbf{q},\omega)
&\equiv -\operatorname{Im}
\left[\epsilon^{-1}(\mathbf{q},\omega)\right]\\
&\approx
\frac{ g_\alpha D_\alpha}{2\pi e^2}
|\mathbf{q}|^{2\alpha-1}
\frac{\omega}
{\omega^2+(D_\alpha|\mathbf{q}|^\alpha)^2}.
\end{aligned}
\end{equation}
   Notably, the relaxational curve in Fig~\ref{fig:EELS}(a) is qualitatively distinct from the loss function peak of underdamped plasmons which are fitted to a Lorentzian in oscillator models~\cite{Goncalves2020Plasmonic}. Our nonlocal model also has a fractional exponent for the dielectric loss peak dispersion, $\omega_* = D_\alpha  q^\alpha$ in Fig~\ref{fig:EELS}(b). In contrast,
   conventional layered models~\cite{Nucker1989}, including those incorporating strong local correlations~\cite{Greco_2016}, can instead yield a quadratic exponent in-plane plasmon dispersion, $\omega_* \sim q^2$~\cite{Greco_2016}.

Placing momentum-resolved electron energy loss spectroscopy (EELS) peak dispersion data of Fig~\ref{fig:EELS}(b) $(q,\omega_*)$ on a log-log plot, one can extract $\alpha_{EELS}$ from the slope, and compare it to $\alpha_{ARPES}$. Our theory sharply predicts a relationship between independently extracted 
\begin{equation}\label{eqn:falsify}
    \alpha_{ARPES}  \equiv \frac{1}{2-\gamma_{ARPES}} = \frac{d \ln \omega_*}{d\ln q} \equiv \alpha_{EELS},
\end{equation}
equal exponents at each fixed doping for which $1<\alpha<2$. Further accounting in our model the layered structure of cuprates, the relaxational curve shape is unchanged but the loss peak dispersion is now gapped, $ \omega_* = \omega_0 + Aq^\alpha + Bq^2 +...$, see Supplemental Material. Our formula for $\omega_*$ is derived by expanding in $qd \ll 1$, requiring $q \ll d^{-1} \sim 0.3\text{\AA}^{-1}$~\cite{leggett2012exotic} ($d$ being the $CuO_2$ inter-plane distance), to probe the subtle effect introduced by the $q^\alpha$ term~\footnote{In fact, if we want to consider the entire unit cell of electrostatically-coupled layers, we have $d_{cell} \approx 12\text{\AA} , 15\text{\AA}$ for Bi2201 and Bi2212 respectively.}. Similar to the self-energy Eqn~\ref{eqn:inelscattering}, the loss function as a function of energy $\omega$ at fixed momentum transfer $|\mathbf{q}|$ has only three fitting parameters: $\omega_0, D_\alpha, \alpha$. Thus, our loss function relaxational curve is quite phenomenologically constrained, similar to the conventional underdamped plasmon Lorentzian fit with also three free parameters. 

We suggest the broad mid-infrared peak around $\sim 1 eV$ in Fig~\ref{fig:EELS}(c)~\cite{Nucker1989, Leggett1999} is the appropriate peak to track with changing momentum. The physics of this broad peak in the loss function ignited debate shortly after the discovery of cuprates and remains highly controversial~\cite{abbamonte2024collectivechargeexcitationsstudied}. Measurements nearly four decades ago~\cite{Nucker1989} on Bi2212 were fitted to a `layered-RPA' relation $\omega_* \sim q^2$ at larger $q^2 \gtrsim 0.05\text{\AA}^{-2}$ (equivalently $q \gtrsim 0.22\text{\AA}^{-1}$), but insufficient experimental resolution $\Delta E \sim 0.15 eV, \ \Delta q \sim 0.04 \text{\AA}^{-1}$ left open the possibility that the $\omega_*(q^2)$ is concave downward at small $q \lesssim 0.22\text{\AA}^{-1}$ in Fig~\ref{fig:EELS}(d).  In the past decade, EELS experiments did not prioritize finely tracking the $q \ll 0.3\text{\AA}^{-1}$ region required to falsify this theory~\cite{abbamonte2024collectivechargeexcitationsstudied, devries2026reexaminingstrangemetalcharge, Mitrano_2018, Zhu2015HREELS}. The $\sim 1$eV peak dispersion exponent in Bi2201 and Bi2212 therefore remains undetermined.

\paragraph*{Conclusion} Cross-spectroscopic comparison of scaling exponents predicted by our model's nonlocal mechanism in cuprates Bi2212 and Bi2201, have become accessible due to advancements in high-resolution EELS~\cite{Zhu2015HREELS} and ARPES~\cite{Liu2008LaserARPES}. Such finely-resolved spectroscopies transform our theoretical proposal into an experimental imperative.

Resonant inelastic x-ray scattering (RIXS) could also provide a complementary probe of the $1$eV broad peak dispersion considered in this work, alongside EELS~\cite{RevModPhys.83.705, schuster2026contrastingonetwoparticleexcitations}. Future exploration of materials beyond cuprates, such as nickelates~\cite{chow2025nickelage}, iron-based superconductors~\cite{Fernandes2022, Paglione2010}, and other layered materials with tunable in-plane carrier density, could help determine whether this phenomenology has wider applicability.

We conclude by reiterating the conceptual point that although $\omega/T$ scaling of observables is commonly associated with quantum criticality, spatial nonlocality provides an alternative mechanism.

\section*{acknowledgements}
I thank Hideaki Nishikawa, Lin Er Chow, Kridsanaphong Limtragool, Hiroyuki Yamase, Jorg Fink, Johannes Schultz, Axel Lubk, Falk Roder and the Xingjiang Zhou ARPES and Xuetao Zhu HREELS groups especially Yiwen Chen, Pengfei Ding, Hengrui Dong and Enling Wang at Institute of Physics, Chinese Academy of Sciences for helpful discussions. I thank the Yukawa Institute for Theoretical Physics for financial support and kind hospitality from the `Young International Researcher Invitation Program' from May-August 2026, and stimulating workshops `Frontiers in Nonequilibrium Physics 2026' and `Quantum thermalization, hydrodynamics and gravity'. I am supported by the CQT Young Researcher Career Development Grant and the National Quantum Scholarships Scheme by the Centre for Quantum Technologies in Singapore.

\nocite{Pippard1953,Phillips_2022,Nishikawa2026Microscopic,Kiselev_2021,RevModPhys.57.287}
\bibliographystyle{apsrev4-2}
\bibliography{references}

@article{Keimer2015,
  author  = {Keimer, B. and Kivelson, S. A. and Norman, M. R. and Uchida, S. and Zaanen, J.},
  title   = {From quantum matter to high-temperature superconductivity in copper oxides},
  journal = {Nature},
  volume  = {518},
  number  = {7538},
  pages   = {179--186},
  year    = {2015},
  doi     = {10.1038/nature14165},
  url     = {https://doi.org/10.1038/nature14165}
}

@article{RevModPhys.83.705,
  title = {Resonant inelastic x-ray scattering studies of elementary excitations},
  author = {Ament, Luuk J. P. and van Veenendaal, Michel and Devereaux, Thomas P. and Hill, John P. and van den Brink, Jeroen},
  journal = {Rev. Mod. Phys.},
  volume = {83},
  issue = {2},
  pages = {705--767},
  numpages = {0},
  year = {2011},
  month = {Jun},
  publisher = {American Physical Society},
  doi = {10.1103/RevModPhys.83.705},
  url = {https://link.aps.org/doi/10.1103/RevModPhys.83.705}
}

@article{chow2025nickelage,
  title   = {Nickel Age of High-Temperature Superconductivity},
  author  = {Chow, S. Lin Er and Ariando, A.},
  journal = {Advanced Materials Interfaces},
  volume  = {12},
  number  = {4},
  pages   = {2400717},
  year    = {2025},
  doi     = {10.1002/admi.202400717}
}

@article{Paglione2010,
  author  = {Paglione, Johnpierre and Greene, Richard L.},
  title   = {High-temperature superconductivity in iron-based materials},
  journal = {Nature Physics},
  volume  = {6},
  pages   = {645--658},
  year    = {2010},
  doi     = {10.1038/nphys1759},
  url     = {https://doi.org/10.1038/nphys1759}
}

@article{Fernandes2022,
  author  = {Fernandes, Rafael M. and
             Coldea, Amalia I. and
             Ding, Hong and
             Fisher, Ian R. and
             Hirschfeld, P. J. and
             Kotliar, Gabriel},
  title   = {Iron pnictides and chalcogenides:
             a new paradigm for superconductivity},
  journal = {Nature},
  volume  = {601},
  pages   = {35--44},
  year    = {2022},
  doi     = {10.1038/s41586-021-04073-2}
}

@article{Helton_2010,
   title={Dynamic Scaling in the Susceptibility of the Spin-<mml:math xmlns:mml=“http://www.w3.org/1998/Math/MathML” display=“inline”><mml:mfrac><mml:mn>1</mml:mn><mml:mn>2</mml:mn></mml:mfrac></mml:math>Kagome Lattice Antiferromagnet Herbertsmithite},
   volume={104},
   ISSN={1079-7114},
   url={http://dx.doi.org/10.1103/PhysRevLett.104.147201},
   DOI={10.1103/physrevlett.104.147201},
   number={14},
   journal={Physical Review Letters},
   publisher={American Physical Society (APS)},
   author={Helton, J. S. and Matan, K. and Shores, M. P. and Nytko, E. A. and Bartlett, B. M. and Qiu, Y. and Nocera, D. G. and Lee, Y. S.},
   year={2010},
   month=Apr }

@article{SachdevKeimer2011,
  author       = {Sachdev, Subir and Keimer, Bernhard},
  title        = {Quantum Criticality},
  journal      = {Physics Today},
  volume       = {64},
  number       = {2},
  pages        = {29--35},
  year         = {2011},
  doi          = {10.1063/1.3554314},
  eprint       = {1102.4628},
  archivePrefix= {arXiv},
  primaryClass = {cond-mat.str-el},
  url          = {https://arxiv.org/abs/1102.4628}
}

@article{Fetter1974LayeredElectronGas,
  author  = {Fetter, Alexander L.},
  title   = {Electrodynamics of a layered electron gas. II. Periodic array},
  journal = {Annals of Physics},
  volume  = {88},
  number  = {1},
  pages   = {1--25},
  year    = {1974},
  doi     = {10.1016/0003-4916(74)90397-2}
}

@article{Greco_2016,
   title={Plasmon excitations in layered high-
                    <mml:math xmlns:mml=“http://www.w3.org/1998/Math/MathML”>
                      <mml:msub>
                        <mml:mi>T</mml:mi>
                        <mml:mi>c</mml:mi>
                      </mml:msub>
                    </mml:math>
                    cuprates},
   volume={94},
   ISSN={2469-9969},
   url={http://dx.doi.org/10.1103/PhysRevB.94.075139},
   DOI={10.1103/physrevb.94.075139},
   number={7},
   journal={Physical Review B},
   publisher={American Physical Society (APS)},
   author={Greco, Andrés and Yamase, Hiroyuki and Bejas, Matías},
   year={2016},
   month=Aug }

@misc{schuster2026contrastingonetwoparticleexcitations,
      title={Contrasting one- and two-particle excitations in cuprate strange metals}, 
      author={Falk Röder. Roman Schuster and Martin Knupfer and Axel Lubk and Jörg Fink},
      year={2026},
      eprint={2611.12569},
      archivePrefix={arXiv},
      primaryClass={cond-mat.str-el},
      url={https://arxiv.org/abs/2611.12569}, 
}

@article{Goncalves2020Plasmonic,
  author  = {Gon{\c{c}}alves, Manuel R. and Minassian, Hayk and Melikyan, Armen},
  title   = {Plasmonic resonators: fundamental properties and applications},
  journal = {Journal of Physics D: Applied Physics},
  volume  = {53},
  number  = {44},
  pages   = {443002},
  year    = {2020},
  doi     = {10.1088/1361-6463/ab96e9},
  url     = {https://doi.org/10.1088/1361-6463/ab96e9}
}

@article{Varma_1989,
  title = {Phenomenology of the normal state of Cu-O high-temperature superconductors},
  author = {Varma, C. M. and Littlewood, P. B. and Schmitt-Rink, S. and Abrahams, E. and Ruckenstein, A. E.},
  journal = {Phys. Rev. Lett.},
  volume = {63},
  issue = {18},
  pages = {1996--1999},
  numpages = {0},
  year = {1989},
  month = {Oct},
  publisher = {American Physical Society},
  doi = {10.1103/PhysRevLett.63.1996},
  url = {https://link.aps.org/doi/10.1103/PhysRevLett.63.1996}
}

@article{Leggett1999,
  title = {A ``midinfrared'' scenario for cuprate superconductivity},
  author = {Leggett, A. J.},
  journal = {Proceedings of the National Academy of Sciences},
  volume = {96},
  number = {15},
  pages = {8365--8368},
  year = {1999},
  month = {July},
  doi = {10.1073/pnas.96.15.8365},
  url = {https://www.pnas.org/doi/abs/10.1073/pnas.96.15.8365},
  publisher = {National Academy of Sciences}
}

@article{Landau1957,
  title = {The Theory of a Fermi Liquid},
  author = {Landau, L. D.},
  journal = {Sov. Phys. JETP},
  volume = {3},
  pages = {920--925},
  year = {1957}
}

@article{Landau1959,
  title = {On the Theory of the Fermi Liquid},
  author = {Landau, L. D.},
  journal = {Sov. Phys. JETP},
  volume = {8},
  pages = {70--74},
  year = {1959}
}

@article{RevModPhys.57.287,
  title = {Disordered electronic systems},
  author = {Lee, Patrick A. and Ramakrishnan, T. V.},
  journal = {Rev. Mod. Phys.},
  volume = {57},
  issue = {2},
  pages = {287--337},
  numpages = {0},
  year = {1985},
  month = {Apr},
  publisher = {American Physical Society},
  doi = {10.1103/RevModPhys.57.287},
  url = {https://link.aps.org/doi/10.1103/RevModPhys.57.287}
}

@article{laughlin2000middle,title={The middle way},author={Laughlin, Robert B. and Pines, David and Schmalian, Joerg and Stojkovi{'c}, Branko P. and Wolynes, Peter G.},journal={Proceedings of the National Academy of Sciences},volume={97},number={1},pages={32--37},year={2000},publisher={National Acad Sciences}}

@article{Emery_1987,
  title = {Theory of high-${\mathrm{T}}_{\mathrm{c}}$ superconductivity in oxides},
  author = {Emery, V. J.},
  journal = {Phys. Rev. Lett.},
  volume = {58},
  issue = {26},
  pages = {2794--2797},
  numpages = {0},
  year = {1987},
  month = {Jun},
  publisher = {American Physical Society},
  doi = {10.1103/PhysRevLett.58.2794},
  url = {https://link.aps.org/doi/10.1103/PhysRevLett.58.2794}
}

@article{Kiselev_2021,
   title={Universal superdiffusive modes in charged two dimensional liquids},
   volume={103},
   ISSN={2469-9969},
   url={http://dx.doi.org/10.1103/PhysRevB.103.235116},
   DOI={10.1103/physrevb.103.235116},
   number={23},
   journal={Physical Review B},
   publisher={American Physical Society (APS)},
   author={Kiselev, Egor I.},
   year={2021},
   month=June }

@article{Hartnoll_2015,
   title={Scaling theory of the cuprate strange metals},
   volume={91},
   ISSN={1550-235X},
   url={http://dx.doi.org/10.1103/PhysRevB.91.155126},
   DOI={10.1103/physrevb.91.155126},
   number={15},
   journal={Physical Review B},
   publisher={American Physical Society (APS)},
   author={Hartnoll, Sean A. and Karch, Andreas},
   year={2015},
   month=Apr }

@book{Altland,
  author = {Altland, Alexander and Simons, Ben D},
  publisher = {Cambridge University Press},
  title = {Condensed matter field theory},
  year = 2010
}

@article{Phillips_2022,
   title={Stranger than metals},
   volume={377},
   ISSN={1095-9203},
   url={http://dx.doi.org/10.1126/science.abh4273},
   DOI={10.1126/science.abh4273},
   number={6602},
   journal={Science},
   publisher={American Association for the Advancement of Science (AAAS)},
   author={Phillips, Philip W. and Hussey, Nigel E. and Abbamonte, Peter},
   year={2022},
   month=July }

@techreport{leggett2012exotic,
  title = {Exotic Superconductivity},
  author = {Leggett, Anthony J.},
  institution = {Kyoto University Research Information Repository},
  year = {2012},
  month = {May},
  url = {https://repository.kulib.kyoto-u.ac.jp/items/cca61295-1c0c-4c3f-bb0f-b2c4b34ee863},
  doi = {10.14989/155815}
}

@article{Hartnoll_2022,
   title={<i>Colloquium</i>
: Planckian dissipation in metals},
   volume={94},
   ISSN={1539-0756},
   url={http://dx.doi.org/10.1103/RevModPhys.94.041002},
   DOI={10.1103/revmodphys.94.041002},
   number={4},
   journal={Reviews of Modern Physics},
   publisher={American Physical Society (APS)},
   author={Hartnoll, Sean A. and Mackenzie, Andrew P.},
   year={2022},
   month=Nov }

@article{Zaanen_2024,
  title   = {Lectures on quantum supreme matter},
  author  = {Zaanen, Jan},
  journal = {Advances in Physics},
  year    = {2024},
  volume  = {73},
  number  = {2},
  pages   = {1--321}, 
  doi     = {10.1080/00018732.2024.2369389},
  url     = {https://www.tandfonline.com/doi/abs/10.1080/00018732.2024.2369389}
}

@article{LiebRobinson1972,
  author       = {Lieb, Elliott H. and Robinson, Derek W.},
  title        = {The Finite Group Velocity of Quantum Spin Systems},
  journal = {Communications in Mathematical Physics},
  year         = {1972},
  volume       = {28},
  number       = {3},
  pages        = {251--257},
  doi          = {10.1007/BF01645779},
  publisher    = {Springer}
}

@article{Kuwahara_2020,
   title={Strictly Linear Light Cones in Long-Range Interacting Systems of Arbitrary Dimensions},
   volume={10},
   ISSN={2160-3308},
   url={http://dx.doi.org/10.1103/PhysRevX.10.031010},
   DOI={10.1103/physrevx.10.031010},
   number={3},
   journal={Physical Review X},
   publisher={American Physical Society (APS)},
   author={Kuwahara, Tomotaka and Saito, Keiji},
   year={2020},
   month=July }

@article{Kuwahara_2021,
   title={Absence of Fast Scrambling in Thermodynamically Stable Long-Range Interacting Systems},
   volume={126},
   ISSN={1079-7114},
   url={http://dx.doi.org/10.1103/PhysRevLett.126.030604},
   DOI={10.1103/physrevlett.126.030604},
   number={3},
   journal={Physical Review Letters},
   publisher={American Physical Society (APS)},
   author={Kuwahara, Tomotaka and Saito, Keiji},
   year={2021},
   month=Jan }

@article{Nishikawa_2025,
   title={Energy Diffusion in the Long-Range Interacting Spin Systems},
   volume={135},
   ISSN={1079-7114},
   url={http://dx.doi.org/10.1103/hsbt-c46n},
   DOI={10.1103/hsbt-c46n},
   number={14},
   journal={Physical Review Letters},
   publisher={American Physical Society (APS)},
   author={Nishikawa, Hideaki and Saito, Keiji},
   year={2025},
   month=Oct }

@phdthesis{Nishikawa2026Microscopic,
  author  = {Nishikawa, Hideaki},
  title   = {Microscopic Theory of Transport Phenomena in the Long-Range Interacting Spin Systems},
  school  = {Kyoto University},
  year    = {2026},
  month   = mar,
  doi     = {10.14989/doctor.k26487},
  type    = {Doctoral dissertation}
}

@article{Zaanen1985,
  title = {Band gaps and electronic structure of transition-metal compounds},
  author = {Zaanen, J. and Sawatzky, G. A. and Allen, J. W.},
  journal = {Physical Review Letters},
  volume = {55},
  issue = {4},
  pages = {418--421},
  year = {1985},
  month = {Jul},
  publisher = {American Physical Society},
  doi = {10.1103/PhysRevLett.55.418},
  url = {https://link.aps.org/doi/10.1103/PhysRevLett.55.418}
}

@article{Gu_2017,
   title={Local criticality, diffusion and chaos in generalized Sachdev-Ye-Kitaev models},
   volume={2017},
   ISSN={1029-8479},
   url={http://dx.doi.org/10.1007/JHEP05(2017)125},
   DOI={10.1007/jhep05(2017)125},
   number={5},
   journal={Journal of High Energy Physics},
   publisher={Springer Science and Business Media LLC},
   author={Gu, Yingfei and Qi, Xiao-Liang and Stanford, Douglas},
   year={2017},
   month=May }

@article{Patel_2018,
   title={Magnetotransport in a Model of a Disordered Strange Metal},
   volume={8},
   ISSN={2160-3308},
   url={http://dx.doi.org/10.1103/PhysRevX.8.021049},
   DOI={10.1103/physrevx.8.021049},
   number={2},
   journal={Physical Review X},
   publisher={American Physical Society (APS)},
   author={Patel, Aavishkar A. and McGreevy, John and Arovas, Daniel P. and Sachdev, Subir},
   year={2018},
   month=May }

@article{Pippard1953,
  author    = {A. B. Pippard},
  title     = {An Experimental and Theoretical Study of the Relation Between Magnetic Field and Current in a Superconductor},
  journal   = {Proceedings of the Royal Society of London. Series A. Mathematical and Physical Sciences},
  volume    = {216},
  number    = {1127},
  pages     = {547--568},
  year      = {1953},
  doi       = {10.1098/rspa.1953.0040}
}

@misc{lange2026realizingemerymodeloptical,
      title={Realizing the Emery Model in Optical Lattices for Quantum Simulation of Cuprates and Nickelates}, 
      author={Hannah Lange and Liyang Qiu and Robin Groth and Andreas von Haaren and Luca Muscarella and Titus Franz and Immanuel Bloch and Fabian Grusdt and Philipp M. Preiss and Annabelle Bohrdt},
      year={2026},
      eprint={2603.11037},
      archivePrefix={arXiv},
      primaryClass={cond-mat.quant-gas},
      url={https://arxiv.org/abs/2603.11037}, 
}

@article{Smit_2024,
   title={Momentum-dependent scaling exponents of nodal self-energies measured in strange metal cuprates and modelled using semi-holography},
   volume={15},
   ISSN={2041-1723},
   url={http://dx.doi.org/10.1038/s41467-024-48594-6},
   DOI={10.1038/s41467-024-48594-6},
   number={1},
   journal={Nature Communications},
   publisher={Springer Science and Business Media LLC},
   author={Smit, S. and Mauri, E. and Bawden, L. and Heringa, F. and Gerritsen, F. and van Heumen, E. and Huang, Y. K. and Kondo, T. and Takeuchi, T. and Hussey, N. E. and Allan, M. and Kim, T. K. and Cacho, C. and Krikun, A. and Schalm, K. and Stoof, H.T.C. and Golden, M. S.},
   year={2024},
   month=May }

@misc{reber2015powerlawliquid,
      title={Power Law Liquid - A Unified Form of Low-Energy Nodal Electronic Interactions in Hole Doped Cuprate Superconductors}, 
      author={T. J. Reber and X. Zhou and N. C. Plumb and S. Parham and J. A. Waugh and Y. Cao and Z. Sun and H. Li and Q. Wang and J. S. Wen and Z. J. Xu and G. Gu and Y. Yoshida and H. Eisaki and G. B. Arnold and D. S. Dessau},
      year={2015},
      eprint={1509.01611},
      archivePrefix={arXiv},
      primaryClass={cond-mat.str-el},
      url={https://arxiv.org/abs/1509.01611}, 
}

@misc{abbamonte2024collectivechargeexcitationsstudied,
      title={Collective charge excitations studied by electron energy-loss spectroscopy}, 
      author={Peter Abbamonte and Jörg Fink},
      year={2024},
      eprint={2404.04670},
      archivePrefix={arXiv},
      primaryClass={cond-mat.str-el},
      url={https://arxiv.org/abs/2404.04670}, 
}

@article{Nucker1989,
  title = {Plasmons and interband transitions in ${\mathrm{Bi}}_{2}$${\mathrm{Sr}}_{2}$Ca${\mathrm{Cu}}_{2}$${\mathrm{O}}_{8}$},
  author = {N\"ucker, N. and Romberg, H. and Nakai, S. and Scheerer, B. and Fink, J. and Yan, Y. F. and Zhao, Z. X.},
  journal = {Phys. Rev. B},
  volume = {39},
  issue = {16},
  pages = {12379(R)--12382(R)},
  numpages = {0},
  year = {1989},
  month = {Jun},
  publisher = {American Physical Society},
  doi = {10.1103/PhysRevB.39.12379},
  url = {https://link.aps.org/doi/10.1103/PhysRevB.39.12379}
}

@article{Zhu2015HREELS,
  title = {High resolution electron energy loss spectroscopy with two-dimensional energy and momentum mapping},
  author = {Zhu, Xuetao and Cao, Yanwei and Zhang, Shuyuan and Jia, Xun and Guo, Qinlin and Yang, Fang and Zhu, Linfan and Zhang, Jiandi and Plummer, E. W. and Guo, Jiandong},
  journal = {Review of Scientific Instruments},
  volume = {86},
  number = {8},
  pages = {083902},
  year = {2015},
  publisher = {AIP Publishing},
  doi = {10.1063/1.4928215},
  url = {https://doi.org/10.1063/1.4928215}
}

@article{Liu2008LaserARPES,
  title = {Development of a vacuum ultraviolet laser-based angle-resolved photoemission system with a superhigh energy resolution better than 1 meV},
  author = {Liu, Guodong and Wang, Guiling and Zhu, Yong and Zhang, Hongbo and Zhang, Guochun and Wang, Xiaoyang and Zhou, Yong and Zhang, Wentao and Liu, Haiyun and Zhao, Lin and Meng, Jianqiao and Dong, Xiaoli and Chen, Chuangtian and Xu, Zuyan and Zhou, X. J.},
  journal = {Review of Scientific Instruments},
  volume = {79},
  number = {2},
  pages = {023105},
  year = {2008},
  publisher = {AIP Publishing},
  doi = {10.1063/1.2835901},
  url = {https://doi.org/10.1063/1.2835901}
}

@misc{devries2026reexaminingstrangemetalcharge,
      title={Reexamining the strange metal charge response with transmission inelastic electron scattering}, 
      author={Niels de Vries and Eric Hoglund and Dipanjan Chaudhuri and Sang hyun Bae and Jin Chen and Xuefei Guo and David Balut and Genda Gu and Pinshane Huang and Jordan Hachtel and Peter Abbamonte},
      year={2026},
      eprint={2602.02348},
      archivePrefix={arXiv},
      primaryClass={cond-mat.str-el},
      url={https://arxiv.org/abs/2602.02348}, 
}

@article{Mitrano_2018,
   title={Anomalous density fluctuations in a strange metal},
   volume={115},
   ISSN={1091-6490},
   url={http://dx.doi.org/10.1073/pnas.1721495115},
   DOI={10.1073/pnas.1721495115},
   number={21},
   journal={Proceedings of the National Academy of Sciences},
   publisher={National Academy of Sciences},
   author={Mitrano, M. and Husain, A. A. and Vig, S. and Kogar, A. and Rak, M. S. and Rubeck, S. I. and Schmalian, J. and Uchoa, B. and Schneeloch, J. and Zhong, R. and Gu, G. D. and Abbamonte, P.},
   year={2018},
   month=May, pages={5392–5396} }

@misc{zhou2006angleresolvedphotoemissionspectroscopyelectronic,
      title={Angle-Resolved Photoemission Spectroscopy on Electronic Structure and Electron-Phonon Coupling in Cuprate Superconductors}, 
      author={X. J. Zhou and T. Cuk and T. Devereaux and N. Nagaosa and Z. -X. Shen},
      year={2006},
      eprint={cond-mat/0604284},
      archivePrefix={arXiv},
      primaryClass={cond-mat.str-el},
      url={https://arxiv.org/abs/cond-mat/0604284}, 
}

\section{Supplemental Material}
\paragraph*{Self-energy integral setup}
We explain how Eqn~\ref{eqn:inelscattering} is set up, for the many-body hydrodynamic medium in thermal equilibrium described by temperature $T$ treated within linear response. This can be done mechanically within the Matsubara formalism~\cite{Altland}, we track the physical processes here. The decay rate of a given process has three crucial constituents,
\begin{equation}
\begin{aligned}
\hspace{-1.0em}
\Gamma_{p}
&= \int_{\mathbf{q}} \int^\infty_0 d\Omega \
    (\text{Transition probability}) \\
&\quad \times (\text{Fermi-Dirac factor ensuring Pauli exclusion}) \\
&\quad \times (\text{Energy conservation delta function}).
\end{aligned}
\end{equation}
Let us first consider the transition probability. Define the measurable retarded density response~\footnote{This refers to the polarization of the material in response to the externally applied field.} (aka measurable polarizability, not the same as the irreducible polarizability) of the many-body medium as 
\begin{equation}
\chi_{nn}^{R}(\mathbf{q},t)
=
-i\Theta(t)
\left\langle
\left[
n(\mathbf{q},t),
n(-\mathbf{q},0)
\right]
\right\rangle_{\rho_\beta}.
\end{equation}
For two many-body eigenstates $\ket{a},\ket{b}$ with energies $E_a,E_b$, let $E_{ab}\equiv E_a-E_b$ and $p_b=e^{-\beta E_b}/Z$. From the Heisenberg picture, the dissipative part is
\begin{equation}
-\operatorname{Im}\chi_{nn}^{R}(\mathbf{q},\Omega)
=
\pi
\sum_{a,b}
(p_b-p_a)
\left|
\left\langle a
\middle|
n(-\mathbf{q},0)
\middle|
b
\right\rangle
\right|^2
\delta(\Omega-E_{ab}).
\end{equation}

 Since $\chi^R_{nn} = \Pi^R_{irr}/(1-\Pi^R_{irr} V_{\alpha})$ in linear response theory, the density response is related to the dynamically screened interaction. Let $\langle a| n(-\mathbf{q},0)|b \rangle \equiv n_{ab}$. With the polarization convention used in the Main Text,
\begin{equation}
W^R(\mathbf{q},\Omega)
=
V_\alpha(\mathbf{q})
+
V_\alpha(\mathbf{q})^2
\chi_{nn}^{R}(\mathbf{q},\Omega).
\end{equation}
Since the instantaneous interaction $V_\alpha(\mathbf{q})$ is real, the dissipative part of $W^R$ is therefore
\begin{equation}
\begin{aligned}
-\operatorname{Im}W^R(\mathbf{q},\Omega)
&=
V_\alpha(\mathbf{q})^2
\left[
-\operatorname{Im}\chi_{nn}^{R}(\mathbf{q},\Omega)
\right]
\\
&=
\pi
\sum_{a,b}
(p_b-p_a)|V_\alpha(\mathbf{q})
n_{ab}|^2
\delta(\Omega-E_{ab}).
\end{aligned}
\end{equation}
Thus, $-\operatorname{Im}W^R$ directly gives the $V_\alpha$-weighted transition spectrum through which the fermion exchanges energy and momentum with the medium.

Importantly, we now exploit the above to obtain identities $(p_b - p_a) = (1-e^{-\beta \Omega})p_b = e^{\beta \Omega}(1-e^{-\beta \Omega}) p_a , \ 1/(1-e^{-\beta \Omega}) = 1+n_B(\Omega), \ e^{-\beta \Omega}/(1-e^{-\beta \Omega}) = n_B(\Omega)$. We now obtain relations for the probability of the fermion losing $\Omega$ to the medium as
\begin{equation}
\sum_{a,b}p_b |V_\alpha(\mathbf{q})n_{ab}|^2 \delta(\Omega - E_{ab}) = \frac{1+n_B(\Omega)}{-\pi} ImW^R(\mathbf{q},\Omega)
\end{equation}
and the probability of the fermion gaining $\Omega$ from the medium as
\begin{equation}
\sum_{a,b}p_a |V_\alpha(\mathbf{q})n_{ab}|^2  \delta(\Omega - E_{ab}) = \frac{n_B(\Omega)}{-\pi} ImW^R(\mathbf{q},\Omega).
\end{equation}

Now we account for the Fermi-Dirac factors. Consider the four processes of (a) electron or (b) hole, (i) emission or (ii) absorption, summarized below omitting $\int_{\mathbf{q}} \int^\infty_0 \frac{d\Omega}{2\pi}[-ImW^R(\mathbf{q},\Omega)]$ with their origin explained next.
\begin{align}
\Gamma_e^{(em)}
&\propto
[1-n_F(\xi_{\mathbf{k-q}})]
[1+n_B(\Omega)]
\delta(\omega-\Omega-\xi_{\mathbf{k-q}}),
\nonumber\\
\Gamma_e^{(abs)}
&\propto
[1-n_F(\xi_{\mathbf{k-q}})]
n_B(\Omega)
\delta(\omega+\Omega-\xi_{\mathbf{k-q}}),
\nonumber\\
\Gamma_h^{(em)}
&\propto
n_F(\xi_{\mathbf{k-q}})
n_B(\Omega)
\delta(\omega-\Omega-\xi_{\mathbf{k-q}}),
\nonumber\\
\Gamma_h^{(abs)}
&\propto
n_F(\xi_{\mathbf{k-q}})
[1+n_B(\Omega)]
\delta(\omega+\Omega-\xi_{\mathbf{k-q}}).
\end{align}
(a)(i) The electron loses energy $\Omega$ to the medium, with post-scattered energy $\xi_{\mathbf{k}-\mathbf{q}}$ satisfying $\omega = \xi_{\mathbf{k}-\mathbf{q}} + \Omega$. This demands the energy conservation term $\delta(\omega - \Omega - \xi_{\mathbf{k}-\mathbf{q}})$ in the integrand. The electron's destination state must be unoccupied in order to simultaneously obey the Pauli principle and allow the electron to enter the intermediate state $\xi_{\mathbf{k}-\mathbf{q}}$. Thus, we append the Fermi-Dirac term $[1- n_F(\xi_{\mathbf{k}-\mathbf{q}})]$ to account for the probability of unoccupation.

(a)(ii) We now have $\omega = \xi_{\mathbf{k}-\mathbf{q}} - \Omega$, with energy conservation $\delta(\omega + \Omega - \xi_{\mathbf{k}-\mathbf{q}})$. The Fermi-Dirac term is $[1- n_F(\xi_{\mathbf{k}-\mathbf{q}})]$ as well.

(b)(i) Now we consider a hole losing energy to the hydrodynamic medium, $\omega = \xi_{\mathbf{k}-\mathbf{q}} + \Omega$ to give energy conservation $\delta(\omega  - \Omega - \xi_{\mathbf{k}-\mathbf{q}})$. Our intermediate state is when an electron leaves the state $\xi_{\mathbf{k}-\mathbf{q}}$, meaning we need it to initially occupy the hole's destination state $\xi_{\mathbf{k}-\mathbf{q}}$. Thus the Fermi-Dirac term is just $n_F(\xi_{\mathbf{k}-\mathbf{q}})$.

(b)(ii) Here, $\omega = \xi_{\mathbf{k}-\mathbf{q}} - \Omega$ such that energy conversation requires $\delta(\omega + \Omega - \xi_{\mathbf{k}-\mathbf{q}})$. The Fermi-Dirac term is $n_F( \xi_{\mathbf{k}-\mathbf{q}})$.

Putting these together collecting the terms with same $\delta$-factor, $\Gamma_{inel} \equiv (\Gamma_{em,e} +\Gamma_{em,h}) + (\Gamma_{abs,e} +\Gamma_{abs,h})$ for all allowed intermediate transitions in $(q,\Omega)$ parameter space, we get a simplified form
\begin{equation}
\begin{aligned}
\Gamma_{\mathrm{inel}}
={}&
\int_{\mathbf{q}}\int_0^\infty
d\Omega \,
\big[-\operatorname{Im}W^R(\mathbf{q},\Omega)\big]
\Big\{
\\
&\big[
1+n_B(\Omega)-n_F(\xi_{\mathbf{k}-\mathbf{q}})
\big]
\delta\big(
\omega-\Omega-\xi_{\mathbf{k}-\mathbf{q}}
\big)
\\
&+
\big[
n_B(\Omega)+n_F(\xi_{\mathbf{k}-\mathbf{q}})
\big]
\delta\big(
\omega+\Omega-\xi_{\mathbf{k}-\mathbf{q}}
\big)
\Big\}.
\end{aligned}
\end{equation}
now use $n_F(-\xi)=1-n_F(\xi)$ on first term. On the second term, change variables $\Omega\rightarrow-\Omega$ and use $n_B(-\Omega)=-[1+n_B(\Omega)],
\
\operatorname{Im}W^R(\mathbf q,-\Omega)
=
-\operatorname{Im}W^R(\mathbf q,\Omega).$
The two positive-frequency branches can be combined into
\begin{align}
\Gamma_{inel} (\mathbf{k}, \omega,T)
&=
\int_{\mathbf q}\int^\infty_{-\infty} d\Omega
[-\operatorname{Im}W^{R}(\mathbf q,\Omega)]
\nonumber\\
&\hspace{-3.5em}\times
\left[
n_B(\Omega)+n_F(-\xi_{\mathbf{k-q}})
\right]
\delta(\omega-\Omega-\xi_{\mathbf{k-q}}),
\label{eq:G0W-imaginary}
\end{align}
\paragraph*{Self-energy integral calculation for $\gamma > 1$ }
We now evaluate Eqn~\ref{eqn:inelscattering} at $\mathbf{k} = \mathbf{k}_F$ for an external fermion inserted on the Fermi surface. 
Our approximations to go from Eqn~\ref{eqn:inelscattering} to Eqn~\ref{eqn:scalecovariant} are:

(A) \textit{Small-Angle Scattering $q \ll k_F$:} When $|\Omega| \gg \max\{ \omega, T \}$, the sum of Bose and Fermi-Dirac factors in Eqn~\ref{eqn:inelscattering} are highly suppressed. Thus, a fermion near the Fermi surface mainly scatters by exchanging energies $|\Omega| \lesssim \max\{ \omega, T \}$. We require the characteristic momentum transferred $q_\Omega \sim (|\Omega|/D_\alpha)^{1/\alpha} \sim (\max\{ \omega, T \}/D_\alpha)^{1/\alpha} \ll k_F$ for small-angle scattering. The last inequality determines the energies $\omega$ for which our theory would be best described by experimental data. Further ignoring curvature corrections $\frac{\partial^2 \xi}{\partial k_i \partial k_j}|_{\mathbf{k}_F}$, the fermion dispersion is linearized, $\xi_{\mathbf{k}_F-\mathbf{q}} \approx \xi_{\mathbf{k}_F} - \mathbf{v_F} \cdot \mathbf{q} = \xi_{\mathbf{k}_F} -v_F q_{\perp}$.

(B) \textit{Tangential Momentum Dominance $|\mathbf{q}| \approx |q_{\parallel}|$:} Split $\int_{\mathbf{q}} = (1/2\pi)^2\int dq_\perp \int dq_\parallel$ relative to the Fermi surface. Note the energy conservation constraint $\delta(\omega - \Omega - \xi_{\mathbf{k}_F - \mathbf{q}})$ gives $ q_{\perp} = (\Omega - \omega)/v_F$. For $\alpha>1$, when $\Omega$ and $\Omega-\omega$ have similar low-energy order, the characteristic momentum satisfies $|q_{\Omega}|/|q_{\perp}|\sim|\Omega|^{(1/\alpha)-1}\rightarrow\infty$, so $|\mathbf q|\approx|q_{\parallel}|$, and $W(\mathbf{q},\Omega) \approx W(q_\parallel,\Omega)$.

Applying the above, the linearization gives a new energy conservation $\delta\!\left(
\omega-\Omega+v_{F}q_{\perp}
\right)$ such that the integral over the momentum component normal to the Fermi
surface is $
\int dq_{\perp}\,
\delta\!\left(
\omega-\Omega+v_{F}q_{\perp}
\right)
=
\frac{1}{v_{F}}$. The remaining momentum variable is the tangential component
$q_{\parallel}$ with a simplified
\begin{equation}
\begin{aligned}
\Gamma_{inel}
&\approx
\frac{1}{4 \pi^2 v_F}
\int^{\infty}_{-\infty} d\Omega
\int^{\infty}_{-\infty} dq_{\parallel}\,
[-\operatorname{Im}W^R(|\mathbf{q}|,\Omega)]
\\
&\qquad\times
\left[
n_B(\Omega)
+
n_F(\Omega-\omega)
\right].
\end{aligned}
\end{equation}
We approximate $|\mathbf{q}| \approx q_{\parallel}$. Now we can exactly evaluate $\int^{\infty}_{-\infty} dq_{\parallel}\,
[-\operatorname{Im}W^R(|\mathbf{q}|,\Omega)] = F_{\alpha}(0) g_{\alpha} D_{\alpha}^{\frac{1}{\alpha}-1} sgn(\Omega)|\Omega|^{1-\frac{1}{\alpha}}$, where $F_\alpha(y) \equiv \int^{\infty}_{-\infty} dx \frac{(x^2 + y^2)^{\alpha - 1}}{1 + (x^2 + y^2)^{\alpha}} , \ F_\alpha(0) = \frac{\pi}{\alpha \sin(\pi/2\alpha)}$. Next, we ignore constants for brevity. We have now reduced to a single integral over $d\Omega$, using $n_B(-\Omega) = -[1+n_B(\Omega)]$ and $1-n_F(-\Omega-\omega) = n_F(\Omega + \omega)$ and splitting the integral back onto $\Omega > 0$,
\begin{equation}
 \Gamma_{inel} \propto
\int^{\infty}_{0} d\Omega \ |\Omega|^{\gamma - 1}
\left[
2n_B(\Omega)
+
n_F(\Omega-\omega) + n_F(\Omega+\omega)
\right]
\end{equation}
Let $\Omega = Ty$ and we have the desired form, mathematically valid for $\gamma, \alpha > 1$.

\paragraph*{Marginal point $\gamma = \alpha = 1$} Taking
$\gamma\rightarrow1^+$ and $\alpha\rightarrow1^+$, we have
$\Gamma_{\mathrm{inel}}(\mathbf{k}_F,\omega,0)=
\frac{g_1}{4\pi v_F}|\omega| \Theta (
\Omega_{\mathrm{UV}}-|\omega|)$, where we put in an ultraviolet
cutoff $\Omega_{\mathrm{UV}}$ by hand. For $|\omega|\ll\Omega_{\mathrm{UV}}$, using the Kramers-Kronig relation our
self-energy is (c.f. Varma's self-energy~\cite{Varma_1989})
\begin{align}
\Sigma^R(\mathbf{k}_F,\omega,0)
&=
-\frac{g_1}{2\pi^2v_F}
\omega
\ln\left(
\frac{\Omega_{\mathrm{UV}}}{|\omega|}
\right)
-i\frac{g_1}{4 \pi v_F}|\omega|.
\end{align}
omitting $O(
\frac{\omega^3}{\Omega_{\mathrm{UV}}^2})$ terms. So $\lim_{\omega\rightarrow0}
\partial_\omega
Re\Sigma^R(\mathbf{k}_F,\omega,0) = -\infty$
and the quasiparticle residue $Z_{{\mathbf{k}}} = [1 - Re \partial_{\omega} \Sigma(\mathbf{k},\omega;T)|_{\omega =E_\mathbf{k} }]^{-1}$ vanishes logarithmically, where $E_{\mathbf{k}} = \xi_{\mathbf{k}} + Re\Sigma^R (\mathbf{k}, E_{\mathbf{k}})$.

\paragraph*{Dielectric loss peak} It is known that there is an anomalously broad peak in the dielectric loss function around $1eV$, which does not vanish even as $q \rightarrow 0$. This is a consequence of the layered structure as we now explain. For a layered system with in-plane momentum $q$ and out-of-plane continuous quasimomentum $q_z$, Fetter showed~\cite{Fetter1974LayeredElectronGas}
\begin{equation}
    V_F(q, q_z) = \frac{2\pi e^2}{\epsilon q} \frac{\sinh(q d)}{\cosh(qd) - \cos(q_{z} d)}
\end{equation}
The main assumption we make here is ordinary $3D$ Coulomb interaction between planes, since the partial $V_\alpha$ screening is an in-plane phenomenon. Then Fourier transform from real space to momentum space for the interaction between two layers $n,m$, and further Fourier transform in layer difference $n-m$. At this stage we only deal with electrostatics, no assumption on the in-plane density response is made at this stage. 

We take our measurable in-plane polarization $\chi^R_{nn}$ as an irreducible building block in the layered construction with respect to the additional $3D$ interaction channel $V_F$, combining them we obtain~\footnote{To rephrase things, the physical assumption we make here is that the anomalous in-plane interaction and the ordinary three-dimensional layered Coulomb field should be regarded as separate sectors.}
\begin{equation}
\chi^R_{nn, layer} ( q , q_z, \omega) = \frac{\chi^R_{nn} (q, \omega)}{1-\chi^R_{nn} (q, \omega) V_F (q , q_z)} .
\end{equation}
For $q d \ll 1, \ q_z = 0$, a direct calculation gives the loss peak,
\begin{equation}
    \omega_* = \frac{4 \pi e^2 \sigma}{\epsilon d} + D_\alpha q^\alpha + (D+\frac{ \pi e^2 \sigma d}{3\epsilon}) q^2 + O(q^4)  
\end{equation}
and besides the gap $\omega_0 = \frac{4\pi e^2\sigma}{\epsilon d}$ we see an $q^\alpha$ term arising, not captured in the RPA layered in-plane plasmon fit used by Nucker et al.~\cite{Nucker1989}. The in-plane loss function is qualitatively identical to the two-dimensional case, under the above approximations it is 
\begin{equation}
L_{layer}(q, q_z = 0, \omega)= \frac{\omega_0  \omega}{ \omega^2 + [\omega_0 + D_\alpha q^\alpha ]^2},
\end{equation}
a relaxational curve.

Both the ideal two-dimensional and the layered model have $FWHM/\omega_* = 2\sqrt{3} \sim 3.5$ independent of $\alpha$. Further considering the fractional dispersion exponent, it seems inappropriate describe the nonlocal model's peak as a plasmon.
\end{document}